\documentclass{article}
\usepackage{spconf_new,amsmath,graphicx}

\usepackage{multirow}

\usepackage{bm}
\usepackage{booktabs}
\usepackage{array}
\usepackage{xcolor}
\usepackage{cleveref}
\usepackage{amssymb}
\usepackage{balance}

\newcommand{\RR}{\mathbb{R}}
\renewcommand{\aa}{\mathbf{a}}
\newcommand{\csv}{\hat{\mathbf{u}}}
\newcommand{\QQ}{I}
\newcommand{\GG}{G}
\newcommand{\ZZ}{Z}

\DeclareMathOperator{\FF}{FF}
\DeclareMathOperator{\LN}{LN}
\DeclareMathOperator{\MHA}{MHA}

\DeclareMathOperator{\LSTM}{LSTM}
\DeclareMathOperator{\PERM}{perm}

\newcolumntype{M}[1]{>{\centering\arraybackslash}m{#1}}

\title{Transformer Attractors for Robust and Efficient End-to-end Neural Diarization}

\name{Lahiru Samarakoon, Samuel J Broughton, Marc H\"ark\"onen, Ivan Fung}
\address{Fano Labs, Hong Kong}
\copyrightnotice{979-8-3503-0689-7/23/\$31.00~\copyright2023 IEEE}
\begin{document}
\maketitle
\begin{abstract}
End-to-end neural diarization with encoder-decoder based attractors (EEND- EDA) is a method to perform diarization in a single neural network. EDA handles the diarization of a flexible number of speakers by using an LSTM-based encoder-decoder that generates a set of speaker-wise attractors in an autoregressive manner. In this paper, we propose to replace EDA with a transformer-based attractor calculation (TA) module. TA is composed of a Combiner block and a Transformer decoder. The main function of the combiner block is to generate conversational dependent (CD) embeddings by incorporating learned conversational information into a global set of embeddings. These CD embeddings will then serve as the input for the transformer decoder. Results on public datasets show that EEND-TA achieves 2.68\% absolute DER improvement over EEND-EDA. EEND-TA inference is 1.28 times faster than that of EEND-EDA.

\end{abstract}
\section{Introduction}
\label{sec:intro}

Speaker diarization systems are utilized to identify different speakers and their speech content in conversations. These conversations are often lengthy, have an unpredictable number of speakers, and include a significant amount of overlapping speech. Consequently, an effective diarization system needs to be able to handle real-life scenarios involving multiple speakers engaged in overlapping conversations \cite{ryant2020third, chung2020spot, kinoshita21_interspeech}

Traditional diarization methods tend to rely on clustering-based approaches. However, such methods are often ill-equipped to handle overlapping segments of speech. This limitation and the reliance on multiple components within a system increases the complexity associated with their deployment and maintenance \cite{chung2020spot, yu2022m2met, anguera2012speaker, park2022review}.

End-to-end neural diarization (EEND) systems are capable of handling overlapping speech by addressing it as a multi-label classification problem \cite{fujita2019end}. Moreover, EEND systems simplify the deployment process as they only require a single model, unlike the traditional approaches that involve multiple components.

By design, the EEND architecture can only predict a predetermined number of speakers, which is usually fixed to be a relatively small number due to the combinatorics of Permutation Invariant Training \cite{yu2017permutation,fujita2019permutation}.
A natural question was whether a single model could predict a potentially unlimited number of speakers.
A first breakthrough came with the introduction of the EEND-EDA model \cite{horiguchi22EENDEDA}.
The Encoder-Decoder Attractor (EDA) module uses an LSTM encoder-decoder architecture to transform the sequence of frames into a variable-length sequence of speaker attractors, which are then used to predict each speaker's speech activity.

Since the advent of EEND-EDA, most improvements have mostly ignored the EDA module.
These include changes to the encoder \cite{horiguchi22EENDEDA,leung2021robust}, loss functions \cite{fujita2023intermediate,leung2021robust}, and improved post-processing techniques \cite{horiguchi21global_local,xue21STB}.
Improvements to the EDA module have mostly been limited to minor tweaks to the LSTM encoder-decoder architecture.
Since the only input to the LSTM decoder is the final state of the encoder, several methods propose injecting more context into the decoder. For example, Pan et al.\ \cite{pan2022towards} propose a Bahdanau-style attention mechanism \cite{bahdanau2014attention} to construct inputs for the LSTM decoder. Similarly, Broughton et al.\ \cite{broughton2023improving} propose constructing inputs to the LSTM decoder by using a learned conversational summary vector.

This work presents EEND-TA, a transformer-based method to replace the LSTM-based EDA module of EEND-EDA.
This follows recent trends of replacing standard neural network layers by attention mechanisms, as evidenced by the widespread adoption of the Transformer architecture \cite{transformer} in a wide range of fields, such as natural language processing \cite{devlin2018bert,GPT,T5}, computer vision \cite{vision_transformer}, reinforcement learning \cite{decision_transformer}, and automatic speech recognition \cite{gulati2020conformer, whisper}.

The proposed Transformer Attractors (TA) module is composed of a Combiner block and a Transformer decoder. The main function of the combiner block is to generate conversationally dependent (CD) embeddings by incorporating learned conversational information into a global set of embeddings. We then use these CD embeddings as the input for the transformer decoder. Conversational information is gathered throughout the encoder layers, as proposed in \cite{broughton2023improving}. We explore several functions for the combiner block. Our experimental results on publicly available real conversations demonstrate that EEND-TA significantly outperforms the current state-of-the-art.

There have been previous attempts at incorporating attention-based layers in neural diarization. In \cite{rybicka2022end}, authors propose to use clustered embeddings as input to a transformer decoder to generate attractors for each speaker. This puts a lot of responsibility on the clustering method to find the correct number of speakers and provide reasonable cluster centroids. It is unclear whether this approach would work in practice, as all experiments are conducted with a speaker counting oracle on simulated data. Perhaps closest to our setup is the one presented in \cite{fujita2023intermediate}.
Instead of using a transformer decoder, authors use a single cross-attention layer to generate a fixed number $N$ of attractors. This makes the model more akin to an EEND model as opposed to EEND-EDA, as it is missing a speaker counting component. Furthermore, their improvements over the EEND-EDA baseline are marginal. In contrast, our method is truly end-to-end, trained on the same objectives as EEND-EDA, and shows noticeable improvement over EEND-EDA on real-life datasets. This is all without any additional ``bells and whistles'', such as iterative refinement \cite{rybicka2022end}, intermediate attractors loss or self conditioning \cite{fujita2023intermediate}, each of which could potentially further boost the performance of our model.

The remainder of the paper is structured as follows: Section \ref{sec:background} provides a brief review of the related work. Section \ref{sec:eend-ta} introduces the proposed EEND-TA method. Section \ref{sec:exp} outlines the experimental setup and Section \ref{sec:results} presents the results. We conclude our work in Section \ref{sec:conclusion}.

\section{Background}
\label{sec:background}

In this section, we review the EEND-EDA model and give a brief introduction to the conversational summary vector (CSV) enhancement for the EEND-EDA. 

\subsection{EEND-EDA}
\label{ssec:eend-eda}

EEND models are the current state-of-the-art for end-to-end diarization \cite{horiguchi22EENDEDA, kinoshita21_interspeech, ueda2022eend, yu2022auxiliary}. EEND-EDA is one such architecture that outperforms conventional EEND methods and allows for diarization of an arbitrary number of speakers \cite{horiguchi22EENDEDA, horiguchi2021hitachi, leung2021end, leung2021robust}. 

Given an acoustic input feature sequence $X \in \mathbb{R}^{D' \times T}$, EEND-EDA simultaneously estimates the likelihood of speech activity for multiple speakers. Here, $X$ is a log-scaled Mel-filterbank feature sequence with dimension $D'$ and length $T$. $X$ is first encoded to produce hidden embedding $E \in \mathbb{R}^{D \times T}$ by a Transformer-based network \cite{transformer, gulati2020conformer}. $E$ is then shuffled in the $T$ dimension for input to an LSTM-based EDA module that predicts speaker-wise attractors $A \in \mathbb{R}^{D \times (S + 1)}$, where $S$ is the number of speakers. 

Predictions $(\textbf{p}_t)^T_{t=1}$ for multiple speakers' speech activities are estimated by a sigmoid function acting on the matrix product of $A$ and $E$. At frame $t$, $\mathbf{p}_t := [p_{t, 1}, ..., p_{t, S}] \in (0, 1)^S$ are the predicted probabilities for $S$ speakers' speech activities.

Diarization loss is computed during training with a permutation invariant objective \cite{yu2017permutation}, by taking the minimum loss for all possible speaker assignments between the predicted probabilities and ground truth labels:
\begin{equation}
    \label{eq:diar}
    \mathcal{L}_\mathrm{diar} = \frac{1}{TS} \min_{i \text{ } \in \text{ perm}(1, ..., S)} \sum^T_{t=1} H(\mathbf{p}^i_t, \mathbf{y}_t),
\end{equation}
where $\PERM(1, \dotsc S)$ is the symmetric group over $S$ speakers, and $\mathbf{p}^i_t$ are the permuted predictions at frame $t$ for any $i \in \PERM(1,\dotsc,S)$. Ground truth labels are denoted by $\mathbf{y}_t := [y_{t, 1}, ..., y_{t, S}] \in \{0, 1\}^S$, where $y_{s, t} = 1$ and $y_{s, t} = 0$ denotes whether speaker $s$ is active or not at frame $t$. The function $H(\cdot, \cdot)$ denotes binary cross entropy.

The EDA module consists of an encoder $\LSTM^{\text{enc}}$ and a decoder $\LSTM^{\text{dec}}$. The final hidden and cell states of the encoder are used to initialize the hidden and cell states of the decoder. At each iteration $s = 1,\dotsc,S+1$, $\LSTM^{\text{dec}}$ takes an input zero vector and generates a speaker-wise attractor:
\begin{align}
    \label{eq:orig_lstmd}
    \textbf{h}^{\text{dec}}_{s}, \textbf{c}^{\text{dec}}_{s} &= \LSTM^{\text{dec}}(\textbf{0}, \textbf{h}^{\text{dec}}_{s-1}, \textbf{c}^{\text{dec}}_{s}),
\end{align}
where attractor $\aa_s \in \RR^D$ for speaker $s$ corresponds to the hidden state $\textbf{h}^{\text{dec}}_{s}$ at that step. 

Speaker count is modeled by a probability vector $q_s = \sigma(\FF(a_s))$, obtained from each attractor $a_s$.
Here $\FF$ is a fully connected feed-forward layer, and $\sigma$ is the sigmoid function.
During inference, we generate a new attractor $a_i$ whenever $q_{i-1} > 0.5$.
Therefore, when training on an $S$ speaker recording, we want that $q_1,\dotsc,q_S = 1$, and $q_{S+1} = 0$.
This translates to the training objective
\begin{align}
   \mathcal{L}_\mathrm{exist} &= \frac{1}{S + 1}H(\mathbf{l}, \textbf{q}),
\end{align}
where $\mathbf{l} := [1,...,1,0]^\top$ is a vector with $S$ ones and 1 zero.

The model is optimized during training with the full objective:
\begin{equation}
    \mathcal{L} = \mathcal{L}_\mathrm{diar} + \lambda \mathcal{L}_\mathrm{exist},
\end{equation}
where $\lambda$ is a hyperparameter.

\subsection{Conversational Summary Vector (CSV) for EEND-EDA}
\label{ssec:summary_vector}

One of the limitations of the EEND-EDA is that the LSTM decoder fully relies on the LSTM encoder to create a summary of the entire recording as in the form of hidden and cell states. The main idea behind the CSV is to enhance the information flow to the LSTM decoder by providing an additional summary.

Drawing on inspiration from the BERT classification token \texttt{[CLS]} in natural language processing \cite{devlin2018bert}, a method to generate a CSV for EEND-EDA was recently suggested \cite{broughton2023improving}. Here, a CSV is learned in the encoder similar to BERT, and is seen by the $\LSTM^{\text{dec}}$ during attractor generation.  This was shown to outperform conventional EEND-EDA methods.

The Conformer encoder responsible for generating output embedding $E$ is modified to additionally output a CSV $\hat{\textbf{u}}$ from the original input feature sequence:
\begin{equation}
    \hat{\textbf{u}}, E = \text{enc}(X),
\end{equation}
where  $\hat{\textbf{u}} \in \mathbb{R}^{D \times 1}$ and $E \in \mathbb{R}^{D \times T}$. Zero vectors input to $\LSTM^{\text{dec}}$, as shown in equation \ref{eq:orig_lstmd}, are replaced with $\hat{\textbf{u}}$ such that each speaker-wise attractor is conditioned on the CSV:
\begin{align}
    \textbf{h}^{\text{dec}}_{s}, \textbf{c}^{\text{dec}}_{s} &= \LSTM^{\text{dec}}(\hat{\textbf{u}}, \textbf{h}^{\text{dec}}_{s-1}, \textbf{c}^{\text{dec}}_{s}),
\end{align}
for $s = 1,\dotsc,S+1$.

The CSV is constructed in a modified Conformer encoder by prepending a learnable special token to the feature sequence $X$. Thus the first frame of every input sequence to the encoder corresponds to the CSV, and the special token has its embedding vector updated by the training objective. To maintain a global representation, the CSV bypasses convolutional modules of the Conformer encoder.

In \cite{broughton2023improving}, authors show that incorporating a CSV to EEND-EDA delivers superior results. Most of the improvements come from recordings with four or more speakers. Furthermore, when using CSV, one can train an EEND-EDA model more robustly with increasing lengths of input utterances. 
This suggests that the vanilla EEND-EDA model has a tendency of forgetting certain speakers, and that the LSTM encoder alone is not enough to capture a rich enough representation of the input recording.
The addition of the CSV as an input to the LSTM decoder alleviates that issue, as it injects context learned inside the Conformer encoder, bypassing the LSTM encoder.

\begin{figure}[t]
  \centerline{\includegraphics[width=9.0cm]{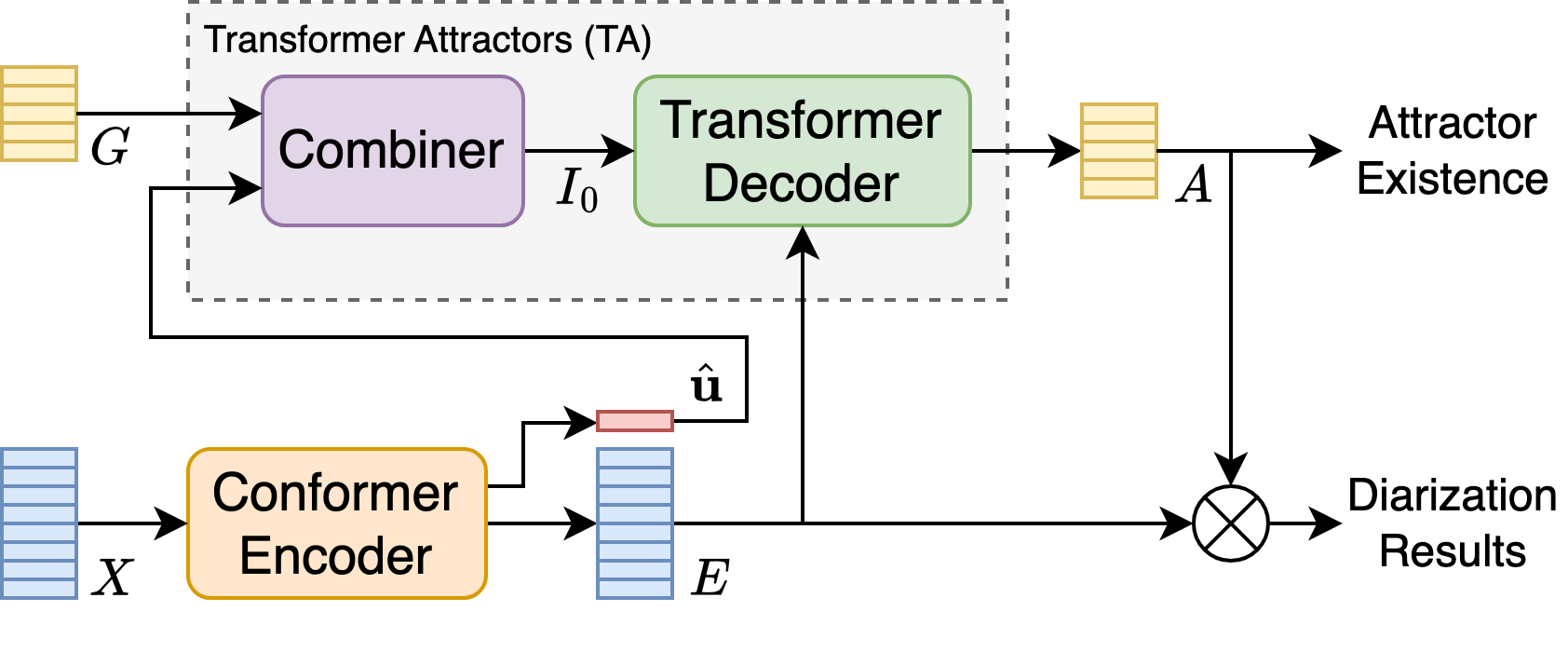}}
  \caption{Illustration of the EEND-TA high-level architecture. EEND-TA contains a Conformer encoder and Transformer Attractors (TA) module. TA consists of a Combiner and a Transformer Decoder.}
  \label{fig:architecture}
\end{figure}

\pagebreak
\section{EEND-TA : Transformer Attractors for EEND}
\label{sec:eend-ta}

In this work, we introduce an EEND model that is free from LSTMs. Concretely, we propose  replacing the LSTM-based EDA with a module called Transformer Attractors (TA) that is based on transformers. The TA module consists of a Combiner block and a Transformer decoder. The role of the combiner block is to prepare a set of inputs for the transformer decoder. This is done by injecting conversational information gathered in the Conformer encoder into a learnable global set of embeddings.
The inputs are then fed to the transformer decoder, using cross-attention with the framewise speech embeddings to output speaker-wise attractors.

Here we present the architecture in detail. The architecture of EEND-TA is depicted in Figure \ref{fig:architecture}. The TA module takes as an input framewise speaker embeddings $E \in \RR^{D \times T}$ and a CSV $\csv \in \RR^{D \times 1}$, and outputs $S+1$ attractors $A = (\aa_i)_{i=1}^{S+1}$, and $S+1$ speaker existence probabilities $q_1,\dotsc,q_{S+1}$. As in EEND-EDA, if there are $S'$ speakers in the conversation, we keep only the first $S'$ attractors and $S'+1$ existence probabilities in the loss computations during training. During inference, we estimate the number of speakers $\hat S$, and output the first $\hat S$ attractors $\aa_1,\dotsc,\aa_{\hat S}$.

\subsection{Combiner}

As show in Figure \ref{fig:architecture}, the transformer decoder's first layer receives inputs $\QQ_0$ that are computed using a combiner function. This function takes the CSV $\csv$ and the learnable global embeddings $\GG \in \mathbb{R}^{D \times (S+1)}$ as inputs. Its purpose is to transform a set of global embeddings into conversationally dependent embeddings. We conduct experiments with various combiner functions.

The first combiner function is
\begin{align}
\label{eqn:add}
\phi_\text{add}(\csv, \GG) = \csv + \GG,
\end{align} where the CSV is added to each global embedding.

Another intuitive function is the element-wise multiplication between CSV and global embeddings
\begin{align}
\phi_\text{mult}(\csv, \GG) = \csv * \GG,
\end{align} where $*$ denotes element-wise multiplication.

Similar to some well-known speaker adaptation mechanisms \cite{swietojanski2016learning, samarakoon2016subspace}, we can learn conversational contributions for global embeddings by treating $\csv$ as a gating function for the entries of $G$:
\begin{align}
\phi_\text{amp}(\csv, \GG) = \alpha \cdot \sigma(\csv) * \GG,
\end{align} where $\alpha > 0$ is the amplitude parameter and $\sigma$ is the sigmoid function.

Finally, to evaluate the importance of conversation-dependent inputs to the transformer decoder, we experiment with removing the combiner altogether, therefore ignoring the conversational summary vector.
Results for these experiments can be found in Table~\ref{tab:comb_func}.

\subsection{Transformer decoder}
We use a standard transformer decoder architecture \cite{transformer} with $N_D$ blocks and no positional encoding.
Each block is fed a set of $S+1$ inputs $\QQ_i \in \RR^{D \times (S+1)}$, which --- together with the encoder output $E \in \RR^{D \times T}$ --- produce the $S+1$ attractors $A \in \RR^{D \times (S+1)}$.
To summarize, each transformer decoder block performs the following computation
\begin{align}
\begin{split}
\ZZ &= \LN(\MHA(\QQ_i, \QQ_i, \QQ_i) + \QQ_i)\\
\ZZ' &= \LN(\MHA(\ZZ, E, E) + \ZZ) \\
\QQ_{i+1} &= \LN(\FF(\ZZ') + \ZZ'),
\end{split}
\end{align}
where $\MHA(Q, K, V)$ is a multi-head attention layer with query $Q$, key $K$, value $V$, $\LN$ is Layer Norm \cite{ba2016layer}, and $\FF$ is a feedforward network.
The output attractors $\aa_1,\dotsc,\aa_{S+1}$ are obtained from the columns of $\QQ_{N_D}$.
Speaker existence probabilities are computed as $q_i = \sigma(\FF(a_i)) \in (0,1)$, where is the sigmoid function.

\subsection{Discussion}

We posit that the transformer architecture is a more natural choice for attractor generation tasks than the LSTM-based EDA. Since LSTM layers are inherently order dependent, special care needs to be taken when presenting the embeddings to the EDA module. The EDA module performs much better when the framewise embeddings are input in shuffled order \cite{horiguchi22EENDEDA}. We hypothesize that this is due to the LSTM cell forgetting speakers appearing earlier in the sequence, and shuffling makes all speakers appear more uniformly throughout the sequence. The transformer decoder overcomes these limitations because it takes into account all framewise embeddings simultaneously and remains unaffected by their order. There are also computational benefits, as the transformer decoder generates all attractors in parallel, compared to the LSTM's sequential frame encoding and speaker decoding, see \Cref{tab:speedup}.

We note that contrary to models in the EEND-EDA family, the maximum number of speakers that EEND-TA can predict is fixed to $S$. While this might seem like a limitation, in practice the situation is slightly more nuanced. Despite being able to predict an unlimited number of speakers, EEND-EDA is empirically shown to be able to predict at most the number of speakers seen in training, with diarization performance dropping dramatically when predicting higher numbers \cite{horiguchi22EENDEDA}.
Thus, due to the scarcity of diarization training data for a high number of speakers, EEND-EDA based models are essentially capped to a certain number of speakers.
This is further evidenced by the use of other techniques such as iterative inference \cite{horiguchi22EENDEDA, xue21STB} or clustering \cite{horiguchi21global_local} when the number of speaker grows too large. These techniques can be readily used with EEND-TA when inferencing for a higher number of speakers.

\section{Experiments}
\label{sec:exp}

\subsection{Datasets}

We use mixtures from the LibriSpeech corpus \cite{panayotov2015librispeech} for all the pre-training phases. These mixtures are simulated using Algorithm 1 outlined in \cite{horiguchi22EENDEDA}. Models are first pre-trained with 100,000 simulated mixtures of 2 speakers with average interval parameter $\beta=2$.  This $\beta$ is a parameter that controls the silence duration and overlap ratio \cite{horiguchi22EENDEDA}. The second phase of the pre-training uses 400,000 simulated mixtures of 1,2,3,4 speakers with $\beta=2,2,5,9$, respectively.

\begin{table}[t]
\caption{Summary of datasets used during finetuning and inference.}
\label{tab:datasets}
\resizebox{\linewidth}{!}{%
    \begin{tabular}{@{}M{3.7cm}M{1.5cm}M{1.5cm}M{1.5cm}@{}}
    \toprule
    \textbf{Dataset} & \textbf{No. of speakers} & \textbf{Finetuning Files} & \textbf{Inference Files} \\ \midrule
        \begin{tabular}[c]{@{}c@{}}
            DIHARD III\\
            VoxConverse\\
            MagicData-RAMC\\
            AMI MIX\\
            AMI SDM1
        \end{tabular}
        & \begin{tabular}[c]{@{}c@{}}
            1 - 10\\
            1 - 21\\
            2\\
            3 - 5\\
            3 - 5
        \end{tabular}
        & \begin{tabular}[c]{@{}c@{}}
            203\\
            216\\
            289\\
            136\\
            135
        \end{tabular} 
        & \begin{tabular}[c]{@{}c@{}}
            259\\
            232\\
            43\\
            16\\
            16\\
        \end{tabular} \\
        \bottomrule
    \end{tabular}
}
\end{table}

During the fine-tuning stage, we use a number of real public datasets: DIHARD III \cite{ryant2020thirdeval}, VoxConverse \cite{chung2020spot}, MagicData-RAMC \cite{yang2022open}, AMI MIX \cite{carletta2006ami} and AMI SDM1 \cite{carletta2006ami}. This creates a combined training set with a total number of 979 files. For our evaluations, test sets from each of the above corpora are used. The details of datasets used during fine-tuning and inference are summarized  in Table \ref{tab:datasets}. 

\subsection{Experimental Setup}

Our EEND-TA and EEND-EDA models consist of 4 Conformer \cite{gulati2020conformer} blocks with 256 hidden units. 
Each Conformer block has four attention heads and feed-forward layers with 1024 hidden units.
No positional encoding is used. The input features are 23-dim log Mel-filterbanks extracted with a window size of 25 ms and a window shift of 10 ms. Unless specified, the input length of the training utterances are set to 5000 frames.
Throughout we have $D = 256$ and $S = 4$, i.e. our model can predict at most 4 speakers.

All the models are first pre-trained with 2-speaker simulated mixtures for 100 epochs, followed by mixtures up to 4 speakers for 25 epochs. The Noam scheduler \cite{transformer} with 100,000 warm-up steps are used during pre-training. During fine-tuning, models are updated for 1000 epochs. We use a fixed learning rate of $5\cdot 10^{-5}$ with the Adam optimizer \cite{kingma2014adam}. The gradient cutting for attractor existence loss is applied before the EDA to avoid back-propagation to the preceding Conformer encoder. All the models are trained to output four dominant speakers due to time complexity of the permutation-invariant loss (PIT) \cite{horiguchi22EENDEDA}.

We evaluate all the models with a diarization threshold of 0.5 and attractor existence threshold of 0.5. The reported DER is based on a collar of 0.0. To obtain the inference model, we average the 10-best checkpoints based on validation DER.

\section{Results}
\label{sec:results}

\begin{table}[htb]
\caption{ Diarization Error Rate (DER) \% for various EEND baselines across all test datasets combined.}
\label{tab:baselines}
\centering
\resizebox{\linewidth}{!}{%
\begin{tabular}{@{}lccccc@{}}
\toprule
\textbf{Model} & \textbf{NS1}   & \textbf{NS2}   & \textbf{NS3}   & \textbf{NS4} & \textbf{NS1 to NS4} \\ \midrule
EDA            & 7.81           & 12.83          & 19.67                & 27.21    & 17.45               \\
EDA + CSV     & 10.37          & 12.78          & 20.35                & 25.62     & 17.13               \\
 \bottomrule
\end{tabular}
}
\end{table}

In Table \ref{tab:baselines}, we report the results for EEND-EDA baseline and EEND-EDA with CSV.  EEND-EDA performance is significantly better than the EEND-EDA with CSV for single speaker test recordings (NS1). However, EEND-EDA with CSV outperforms the EEND-EDA baseline by 0.32\% absolute DER improvement for recordings from one to four active speakers (NS1 to NS4).
Most of the gains of EDA with CSV are coming from four active speaker test recordings (NS4) which is a 1.59\% absolute DER improvement. This shows that CSV enhancement to EDA facilitates remembering more speakers.

\begin{table}[htb]
\caption{ DER (\%) for different numbers of Transformer Decoder layers.}
\label{tab:num_blocks}
\centering
\resizebox{\linewidth}{!}{%
\begin{tabular}{@{}cccccc@{}}
\toprule
\textbf{\# Layers} & \textbf{NS1}   & \textbf{NS2}   & \textbf{NS3}   & \textbf{NS4} & \textbf{NS1 to NS4} \\ \midrule
1     & 7.13           & 11.87          & 17.29                & 25.98    & 16.30               \\
2     & 7.75          & 11.82          & 18.55                & 24.34     & 15.93               \\
3     & 8.61         & 11.47          & 17.80                & 22.98     & 15.30               \\
4     & 8.72          & 11.55          & 18.90                & 22.83     & 15.40               \\
 \bottomrule
\end{tabular}
}
\end{table}

Next, we conduct experiments using EEND-TA. Specifically, we investigate the number of decoder layers for the Transformer decoder. In all the experiments mentioned in Table \ref{tab:num_blocks}, we utilize the combiner function $\phi_\text{add}$ as stated in Equation \ref{eqn:add}. It shows that the optimal performance (15.30\%) is achieved when we employ 3 decoder layers. When compared to the EEND-EDA baseline, there is a 2.15\% absolute DER improvement. Moreover, the EEND-TA with 3 decoder layers performs better than the EEND-EDA with CSV, with a 1.83\% absolute DER improvement. Therefore, for the remainder of our experiments, we will use 3 decoder layers with EEND-TA.

\begin{table}[htb]
\caption{ DER (\%) for various Combiner functions. The function ``None'' means the combiner is skipped. }
\label{tab:comb_func}
\centering
\resizebox{\linewidth}{!}{%
\begin{tabular}{@{}lccccc@{}}
\toprule
\textbf{Combiner Func.} & \textbf{NS1}   & \textbf{NS2}   & \textbf{NS3}   & \textbf{NS4} & \textbf{NS1 to NS4} \\ \midrule
None     & 8.59         & 11.40          & 17.69                & 23.67     & 15.46               \\
$\phi_\text{add}$      & 8.61         & 11.47          & 17.80                & 22.98     & 15.30               \\
$\phi_\text{multi}$    & 7.51         & 11.40          & 17.97                & 23.83     & 15.48               \\
$\phi_\text{amp},_{\alpha=0.8}$    & 8.66          & 11.55          & 17.64                & 22.92     & 15.31               \\
$\phi_\text{amp},_{\alpha=1.0}$     & 8.87          & 11.42          & 17.76                & 21.25     & 14.77               \\
$\phi_\text{amp},_{\alpha=3.0}$     & 7.96          & 11.43          & 18.61                & 24.02     & 15.62               \\
$\phi_\text{amp},_{\alpha=5.0}$    & 8.48          & 11.40          & 18.16                & 22.80     & 15.23               \\
$\phi_\text{amp},_{\alpha=10.0}$     & 8.00          & 11.47          & 17.28                & 22.47     & 15.07               \\
$\phi_\text{amp},_{\alpha=15.0}$    & 8.75          & 11.41          & 17.88                & 23.25     & 15.36               \\
 \bottomrule
\end{tabular}
}
\end{table}

We present the results for all the combining functions in Table \ref{tab:comb_func}. The best performance is achieved with the $\phi_\text{amp},_{\alpha=1.0}$ function. That combiner function will be used for all subsequent experiments.
We also study the significance of incorporating the CSV into the inputs to the transformer decoder by comparing with a model where the CSV is ignored. We observe a decline in performance when ignoring the CSV for the NS4 test sets, with the error rate increasing to 23.67\%, higher than almost every other combiner function. This highlights the importance of conversational summaries for recordings that involve more than just a few speakers. These findings align with the conclusions drawn in \cite{broughton2023improving}.

\begin{table}[htb]
\caption{ Summary of the results for the best models for each technique.}
\label{tab:summary}
\centering
\begin{tabular}{@{}lcc@{}}
\toprule
\textbf{Attractor}  & \textbf{NS1 to NS4} & \textbf{NS1 to NS9} \\ \midrule
EDA                    & 17.45     &    21.68           \\
EDA + CSV                    & 17.13     &  21.34              \\
TA                   & 14.77     &    18.78            \\
 \bottomrule
\end{tabular}
\end{table}

In Table \ref{tab:summary}, we summarize the results for each model. Clearly EEND-TA outperforms by a significant margin both EEND-EDA and EEND-EDA with CSV. When considering the combined test sets from NS1 to NS4, EEND-TA demonstrates absolute DER improvements of 2.68\% and 2.36\% over EEND-EDA and EEND-EDA with CSV, respectively.

For completeness, we also present the results for the combined test sets up to 9 speakers.
For EEND-EDA (+ CSV), we let the model predict as many speakers as necessary until the speaker existence probability drops below 50\%.
The EEND-TA model however is fixed to predicting a maximum of 4 speakers.
Remarkably, we observe considerably better overall performance of the EEND-TA, despite having a hard limit on the number of predicted speakers.
This observation brings further evidence to the ``empirical upper limit'' of the EEND-EDA family of models: while the number of speakers predicted is a priori uncapped, it is more beneficial \emph{not} to make the extra speaker prediction if they are of poor quality.

\begin{table}[htb]
\caption{DER (\%) breakdown for each test set with EEND-TA's relative improvements over EEND-EDA.}
\label{tab:all-datasets}
\centering
\resizebox{\linewidth}{!}{%
\begin{tabular}{@{}llcc@{}}
\toprule
\textbf{Dataset}                & \textbf{Attractor} & \textbf{NS1 to NS4}  & \textbf{Rel. Improvement} \\ \midrule
\multirow{2}{*}{DIHARD III}     & EDA       & 14.07               &  -    \\
                                & TA     & 12.93               & 8.10\%               \\ \midrule
\multirow{2}{*}{VoxConverse}    & EDA       &  15.75               &  -              \\
                                & TA     & 9.89                & 37.2\%      \\ \midrule
\multirow{2}{*}{MagicData-RAMC} & EDA       & 14.45                  &  -    \\
                                & TA     & 13.58                 & 6.02\%              \\ \midrule
\multirow{2}{*}{AMI Mix}        & EDA       &  19.85                   & -               \\
                                & TA     & 17.88                  & 9.92\%      \\ \midrule
\multirow{2}{*}{AMI SDM1}       & EDA       & 32.24                    &  -              \\
                                & TA     & 24.64                    & 23.55\%     \\ \bottomrule
\end{tabular}
}
\end{table}

In Table~\ref{tab:all-datasets} we show that EEND-TA shows significant improvement across all test sets, demonstrating the generalizability of our approach. It is worth noting the substantial relative improvements of 37.2\% and 23.55\% on VoxConverse and AMI SDM1 respectively.
The smaller relative gains in the MagicData-RAMC dataset are due to it only consisting of conversations between 2 speakers. In contrast, VoxConverse includes more recordings with higher number of speakers, and EEND-TA shows the highest relative improvement on that dataset.
It is worth noting that the AMI SDM1 data is recorded in far-field conditions. This suggests the possibility that EEND-TA is more robust to far-field conditions compared to EEND-EDA.

\begin{table}[]
\caption{ Comparison of inference throughput of the baselines and EEND-TA.}
\label{tab:speedup}
\centering
\begin{tabular}{@{}lcc@{}}
\toprule
\textbf{Attractor}  & \textbf{Speed Up }  & \textbf{\# Params(M)}\\ \midrule
EDA                    & 1.00x     &   8.1     \\
EDA + CSV                    & 0.94x     & 8.1          \\
TA                   & 1.28x   &        10.2      \\
 \bottomrule
\end{tabular}
\end{table}

Next, we examine the inference efficiency of the EEND-TA compared to the baselines. Table 8 reveals that EEND-TA can process audio 1.28 times faster than EEND-EDA. This behavior is notable considering that EEND-TA has more parameters (10.2 Million) than EEND-EDA (8.1 Million).
We attribute the increase in performance to the TA module being able to predict all attractors in parallel, whereas the LSTM needs to process each frame and each speaker sequentially.
We further note that the addition of the transformer decoder does not increase computational or space complexity.
The framewise self-attention layers in the Conformer encoder each have complexity $\mathcal{O}(T^2)$.
In the transformer decoder, the self-attention layers are only $\mathcal{O}(S^2)$, whereas the cross-attention layers are $\mathcal{O}(TS)$.
Since $T$ is usually orders of magnitude larger than $S$ (in our case $T = 500$, $S = 4$), the computational and space complexity are dominated by the encoder part.

\begin{table}[]
\caption{DER (\%) results when various lengths of input is used during fine-tuning.}
\label{tab:input_lengths}
\centering
\resizebox{\linewidth}{!}{%
\begin{tabular}{@{}lcccccc@{}}
\toprule
\textbf{Model}                & \textbf{Length} & \textbf{NS1}  & \textbf{NS2} & \textbf{NS3} & \textbf{NS4} & \textbf{NS1 to NS4} \\ \midrule
\multirow{4}{*}{EEND-EDA}     & 50s        &  7.81 & 12.83  & 19.67  & 27.21            & 17.45     \\
                              & 100s       &  8.11 &  12.56 &  20.61 &  24.93           & 16.72    \\
                              & 150s       &  7.89 & 13.03  & 19.76  & 25.55            & 17.08    \\
                              & 200s       &  7.59 & 11.88  & 19.41  & 25.51            & 16.37             \\ 
                                \midrule
\multirow{4}{*}{EEND-TA}    & 50s       &  8.87 & 11.42 & 17.76 & 21.25               & 14.77              \\
                            & 100s      &  7.23 & 11.35 & 17.49 & 20.47               & 14.39     \\
                            & 150s      &  9.26 & 11.49 & 16.84 & 20.29               & 14.46     \\
                            & 200s      &  6.68 & 11.44 & 17.87 & 18.91               & 13.99            \\  
                                \bottomrule
\end{tabular}
}
\end{table}

Finally, in Table \ref{tab:input_lengths}, we compare the performance of EEND-TA and EEND-EDA when trained with long utterances. Both EEND-EDA and EEND-TA show consistent improvements as we increase the training length. Specifically, EEND-EDA's performance improves from 17.45\% to 16.37\%. However, even with only 50 seconds of training, EEND-TA outperforms EEND-EDA with 200 seconds of training. For EEND-TA, performance improves from 14.77\% to 13.99\% as we increase the training utterance length. Notably, when we increase the training utterance length, EEND-EDA and EEND-TA report relative improvements over NS4 test sets of 6.25\% and 11.0\%, respectively. This further demonstrates that EEND-TA extracts speakers more accurately. The best EEND-TA model achieves a 2.38\% absolute DER reduction compared to the best EEND-EDA model. Increasing the training utterance length further is impractical due to hardware limitations.

\section{Conclusion}
\label{sec:conclusion}

In this work we propose EEND-TA, a novel end-to-end diarization model.
The model uses a transformer decoder to predict speaker attractors, combining conversationally dependent inputs with the encoded framewise memory in a cross-attention layer.
This is in stark contrast with traditional, LSTM-based EEND-EDA models, whose only memory of the acoustic information is a collapsed representation obtained from the LSTM encoder.
Similarly to the EEND-EDA model, we show that performance of the EEND-TA model can be improved by injecting a conversational summary vector, which aggregates context from deep in the Conformer encoder.
We show that our approach outperforms EEND-EDA, with major improvements in recordings with higher number of speakers and far-field recordings.
EEND-TA also enjoys improved accuracy when trained with longer sequences.
Despite our best model having slightly more parameters than the best EEND-EDA model, it is computationally faster due to the inherently parallel nature of attention mechanisms.

\pagebreak
\bibliographystyle{IEEEbib}
\bibliography{refs}

\end{document}